\documentclass[aps,prd,twocolumn,showpacs,superscriptaddress]{revtex4}
\usepackage{graphicx,amssymb}
\usepackage{flafter}

\let\oldsqrt\sqrt
\def\sqrt{\mathpalette\DHLhksqrt}
\def\DHLhksqrt#1#2{%
\setbox0=\hbox{$#1\oldsqrt{#2\,}$}\dimen0=\ht0
\advance\dimen0-0.2\ht0 
\setbox2=\hbox{\vrule height\ht0 depth -\dimen0}%
{\box0\lower0.4pt\box2}}

\begin{document}
\preprint{Study of  $\psi(2S)$ decays into $\gamma K^+K^-$ and $\gamma \pi^+\pi^-$}

\title{\boldmath Study of $\psi(2S)$ decays into $\gamma K^+K^-$ and $\gamma \pi^+\pi^-$}
\author{M. Ablikim}
\author{J. Z. Bai}
\affiliation{Institute of High Energy Physics, Beijing 100049, People's Republic of China}
\author{Y. Ban}
\affiliation{Peking University, Beijing 100871, People's Republic of China}
\author{X. Cai}
\affiliation{Institute of High Energy Physics, Beijing 100049, People's Republic of China}
\author{H.F. Chen}
\affiliation{University of Science and Technology of China, Hefei 230026, People's Republic of China}
\author{H.S. Chen}
\author{H.X. Chen}
\author{J.C. Chen}
\author{Jin Chen}
\author{Y.B. Chen}
\author{Y.P. Chu}
\affiliation{Institute of High Energy Physics, Beijing 100049, People's Republic of China}
% ...
\collaboration{BES Collaboration}
\noaffiliation
\date{\today}

\begin{abstract}
  Radiative charmonium decays from the BESII sample of
  14$\times10^{6}$ $\psi(2S)$-events into two different final states,
  $\gamma K^+K^-$ and $\gamma\pi^+\pi^-$, are analyzed.  Product
  branching fractions for $\psi(2S)\rightarrow\gamma X\rightarrow
  \gamma\pi^+\pi^-$, $\gamma K^+K^-$ are given, where $X=f_2(1270)$,
  $f_0(1500)$, and $f_0(1710)$ in $\pi^+\pi^-$ and $f_2(1270)$,
  $f_2'(1525)$, and $f_0(1700)$ in $K^+K^-$.  An angular analysis
  gives the ratios of the helicity projections for the $f_2(1270)$ in
  $\psi(2S)\rightarrow\gamma f_2(1270)\rightarrow\gamma\pi^+\pi^-$.
\end{abstract}


\maketitle

\section{Introduction}
Several exotic QCD states, such as glueballs, hybrids or multiquarks,
are expected to have masses between 1 and 2.5 GeV/$c^2$ \cite{PDG}.
Radiative $J/\psi$ and $\psi(2S)$ decays are traditionally regarded as
gluon-rich environments; such decays, although well investigated, are
still promising for the discovery and identification of QCD exotica.
For example, in radiative decays into pseudoscalars, two of the most
promising glueball candidates, the $f_0(1500)$ and the $f_0(1710)$,
are seen \cite{PDG}.  Several high precision partial wave analyses of
radiative $J/\psi$ decays have been performed \cite{YellowReport}, and
a recent BES-paper presented an analysis of a high statistics sample
of $\psi(2S)$ decays, without a partial wave analysis
\cite{psipradiative}.  This paper updates the $\psi(2S)$ radiative
decays \cite{psipradiative, bes1pseudo} with $\gamma K^+K^-$ and
$\gamma\pi^+\pi^-$ final states, giving access to scalar and tensor
intermediate states.  

Within the quark model \cite{charmthemodel}, the $\psi(2S)$
is believed to have 2 $^3S_1$ as its main component, and the $J/\psi$
is the 1 $^3S_1$ $c\bar{c}$-state. This gives a $\psi(2S)$ with a
behavior that differs from $J/\psi$ only due to the difference in
energy scale and in the radial wave function. The so-called 12\%-rule
from perturbative QCD (pQCD) relates the branching fractions of the two
states into a particular hadronic final state ($h.f.s.$):

\begin{eqnarray}
\frac{BR(\psi(2S)\rightarrow h.f.s.)}{BR(J/\psi\rightarrow h.f.s.)} &\stackrel{pQCD}{=} & \frac{BR(\psi(2S)\rightarrow ggg)}{BR(J/\psi\rightarrow ggg)} \\ \nonumber
&\backsimeq&\frac{BR(\psi(2S)\rightarrow e^+e^-)}{BR(J/\psi\rightarrow e^+e^-)} \\ \nonumber
&=&\frac{(7.43\pm0.06)\times 10^{-3}}{(5.94\pm0.06)\times 10^{-2}} \\ \nonumber
&\backsimeq& 12.5\%,
\end{eqnarray}
where the branching fractions are from Ref. \cite{PDG}.  A radiative
decay mode would be similar, but would proceed via one photon and two
gluons, exchanging one power of $\alpha_{s}$ with $\alpha$ in
the coupling
\cite{Appelquist, Chanowitz, Okun, Brodsky, Koller}. 
A large violation of the 12\%-rule was first observed in 1983 
in $\psi(2S)$ decays to $\rho\pi$ and $K^{*+}K^-+c.c.$ by
MARKII; it became known as the $\rho\pi$-puzzle \cite{MARKIIrhopi}.
Since then, many two-body decay modes have been compared, of which some obey
and some violate the
rule 
\cite{PDG, bes3pi}.

The analysis uses the BESII data sample of
$(14\pm0.56)\times10^6$ $\psi(2S)$-events \cite{numberofevents},
corresponding to an integrated luminosity of $(19.72\pm 0.86)$
pb$^{-1}$ \cite{luminosity}, and a set of continuum data at
$\sqrt{s}=3.65$ GeV, ($6.42\pm 0.24$) pb$^{-1}$
\cite{integratedluminosity}, also measured with the BESII detector.


\section{The BESII detector}
BESII is a cylindrical multi-component detector, described in detail in Refs.
\cite{BES1} and \cite{BES2}. Around the beam-pipe is
a 12-layer straw vertex chamber, which 
provides trigger and track information.
Located radially outside, there is a 40-layer open-cell geometry
main drift chamber (MDC), covering 85\% of the solid angle. The MDC
is used for tracking and particle identification using
$dE/dx$-techniques. 
The momentum resolution is
$\sigma_p/p=1.78\%\sqrt{1+p^2~[\mathrm{GeV}/c]}$, and the MDC
$(dE/dx)/E$ resolution for hadron tracks is $\sim$8\%.  Particle
identification by energy loss techniques is complemented by a
measurement of the time-of-flight (TOF) from the interaction point to
48 scintillation counters surrounding the MDC (time resolution
$\sim$200 ps for hadrons).  Outside the TOF-scintillators is a 12
radiation length lead-gas barrel shower counter (BSC), which measures
the energy of electrons and photons over $\sim$80\% of the total solid
angle with a resolution of $\sigma_E/E=22\%/\sqrt{E~[\mathrm{GeV}]}$.
In addition, the iron flux return, outside the solenoid coil (0.4 T
axial field), is instrumented with three double layers of counters for
identification of muons with momenta larger than 0.5 GeV/$c$.  The
cylindrical structure is closed by end-caps.

\section{Event selection}
Events with at least one photon and two oppositely charged tracks with
good helix fits and complete covariance matrices are selected.
End-cap information is used to reject background with extra charged
particles in the end-cap region, but only events within the
well-understood barrel detectors are included in the analysis.  The
polar angles of the charged particles are required to fulfill
$|\cos\theta_{\pm}|<0.8$ for $K^+K^-$ and $|\cos\theta_{\pm}|<0.65$
for $\pi^+\pi^-$, where $\theta_\pm$ is the angle of the charged particle
with respect to the beam axis.  The vertex of the two charged tracks
is required to satisfy $V_{xy} = \sqrt{V_x^2 + V_y^2} < 1.5$ cm and
$|V_z|<15$ cm, where $V_x$, $V_y$, and $V_z$ are the $x$, $y$, and $z$
coordinates of the point of closest approach of each charged track to
the beam axis.  
The photon with the highest energy is tested for
quality: it must be within the barrel detector, it must have its
first hit in the BSC inside the first 12 
layers out of 24, it must be separated from charged tracks by at
least 15$^\circ$ in the xy-plane at the entrance of the BSC,
and the photon direction from the origin to the BSC
must agree within 35$^\circ$ with the shower direction in
the BSC.

\subsection{Particle identification}
The most abundant $e^+e^-$ reactions at GeV-energies are QED
processes, mainly $e^+e^-\rightarrow e^+e^-$ and
$e^+e^-\rightarrow\mu^+\mu^-$. In cases where there is initial or
final state radiation, the event could be misidentified as
$\gamma\pi^+\pi^-$ or $\gamma K^+K^-$. 
Muons are rejected using muon chamber information;
this is especially important in the $\gamma\pi^+\pi^-$-channel,
therefore the stricter angular cut.  Bhabha events,
$e^+e^-\rightarrow\gamma e^+e^-$, are rejected using the $dE/dx$
$\chi$-value under the $e^\pm$ hypothesis 
($\chi_e=(dE/dx_{measured}-dE/dx_{expected\: for\: e^\pm})/\sigma_{dE/dx\: e^\pm}$)
combined with the energy deposited
in the calorimeter $E_{cal}$ and the momentum $p_{MDC}$. 
To
reject background from $e^+e^-\rightarrow (\gamma)e^+e^-$, the
selected sample is required to satisfy
\begin{equation}
\chi_+^2+\chi_-^2+6\left(\sqrt{(\frac{E_{cal.}}{p_{MDC}})^2_+
+(\frac{E_{cal.}}{p_{MDC}})^2_-}-2\right)^2 >5^2. 
\label{eq:ellipsoid1}
\end{equation}

The TOF and $dE/dx$ measurements of the two charged tracks and a
kinematic fit with four degrees of freedom are used to calculate
$\chi^2$-values for the overall event-hypothesis: $\gamma\pi^+\pi^-$
or $\gamma K^+K^-$. A confidence level of better than 1\% is required
for each event sample, and for ambiguous cases, the most likely
assignment is chosen.  After all requirements, the overall detection
and event selection efficiency is $\sim$15\%
(see Table
~\ref{tab:pinumevents} and ~\ref{tab:knumevents} for exact numbers). The
probability that a $\gamma K^+K^-$ event is selected as a $\gamma
\pi^+\pi^-$ event is 0.17\%, and the probability that a $\gamma
\pi^+\pi^-$ event is selected as a $\gamma K^+K^-$ event is 0.14\%, from Monte
Carlo simulation.

\subsection{Background}
\begin{figure}
\begin{center}
\includegraphics[width=\linewidth]{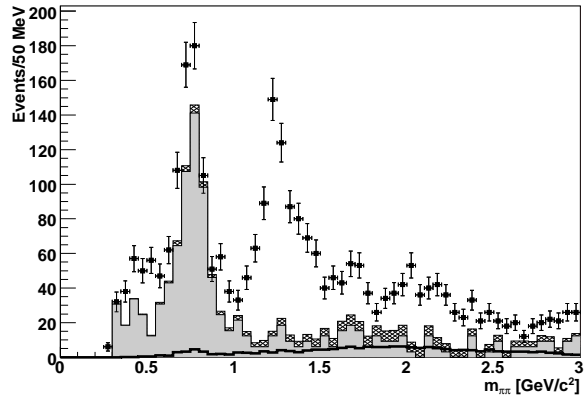}
\caption{The $m_{\pi^+\pi^-}$ distribution for $\psi(2S) \to \gamma\pi^+\pi^-$
candidate events (dots with error bars), along with 
the estimated background.
The thick line corresponds to the background from 
$\psi(2S)\rightarrow\pi^0\pi^-\pi^+$ using the branching fraction from
\cite{PDG}, 
the filled histogram 
corresponds to the continuum data, and the hatched histogram
to a background estimate 
including the continuum and misidentified 
$\psi(2S)\rightarrow\pi^0\pi^-\pi^+$-events. All contributions are scaled
to correspond to the integrated luminosity of the $\psi(2S)$ data set.
The histogram is not acceptance corrected.}
\label{fig:pi}
\end{center}
\end{figure}

\begin{figure}
\begin{center}
\includegraphics[width=\linewidth]{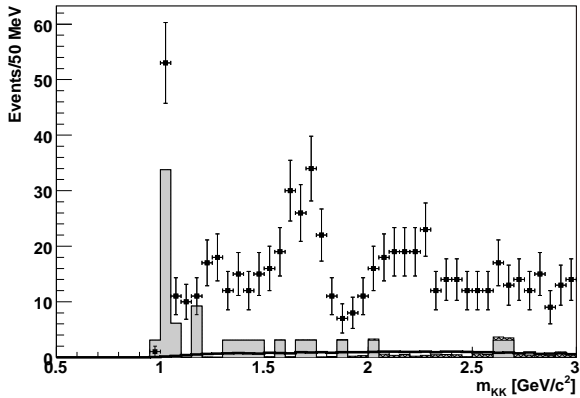}
\caption{The $m_{K^+K^-}$ distribution in $\psi(2S) \to \gamma
  K^+K^-$ candidate events (dots with error bars) along with the
  extimated background. The thick line corresponds to the maximum
  background from $\psi(2S)\rightarrow K^+K^-\pi^0$, which could be
  important if it is close to its upper limit \cite{PDG}.  The filled
  histogram corresponds to the scaled continuum, and the hatched
  histogram to a background estimate including the
  continuum and misidentified
  $\psi(2S)\rightarrow\pi^0\pi^-\pi^+$-events. All contributions are
  scaled to correspond to the integrated luminosity of the $\psi(2S)$
  data set.  The histogram is not acceptance corrected.}
\label{fig:kk08}
\end{center}
\end{figure}

A carefully checked Geant3-based Monte Carlo program \cite{gensim} is
used to determine invariant mass resolution ($<$10 MeV/$c^2$ for
$m_{\pi^+\pi^-}$ and for $m_{K^+K^-}$), angular efficiency, absolute
efficiency, and background suppression. Using this program, the
contributions from QED-processes after event selection are found to be
small (dominated by 15$\pm$2 $\gamma\mu^+\mu^-$ events in
$\gamma\pi^+\pi^-$).  However, both QED backgrounds and non-resonant
hadronic signals are estimated using the continuum data sample, where
the main contributor to the non-resonant background is found to be
initial-state-radiation and subsequent $\rho$ or $\phi$-formation.
This was checked for the $\rho$, and measurement agreed
with calculations using a structure function approach
\cite{YellowReport, Lundborg}.  The full continuum contribution is
assumed to be incoherent with the signal.  Resonant background sources
are checked individually using known cross sections and simulation to determine the acceptance
of the selection criteria \cite{PDG,
  beskkstar, omegapi0}.  The main resonant background to both channels
is found to be $\psi(2S)\rightarrow\pi^0\pi^+\pi^-$, which has
been measured with great precision by BESII \cite{bes3pi} using
partial wave analysis. A Monte Carlo generated sample with the same
angular and energy distribution as found in Ref. \cite{bes3pi} is
used, and 72$\pm$5 misidentified background events in
$\gamma\pi^+\pi^-$ and 4.9$\pm$0.5 misidentified events in $\gamma
K^+K^-$ are obtained.  The signal and the background estimates are
shown in Figs. \ref{fig:pi} and \ref{fig:kk08}.

\section{Intermediate resonances}
From spin-parity conservation, any intermediate resonance $X$
in $e^+e^-\rightarrow \psi(2S)\rightarrow \gamma X\rightarrow \gamma K^+K^-$,
$\gamma\pi^+\pi^-$ must have $J^{PC}=0^{++}$, $2^{++}$, or higher even spins.
To investigate the resonance structure, the invariant mass spectra of the
two pseudoscalar systems are fitted.

\subsection{Previous observations}
A partial wave analysis of BESII $J/\psi\rightarrow \gamma
\pi\bar{\pi}$-data \cite{jpsipwa} required an $f_2(1270)$ at
$m=1262^{+1}_{-2}\pm6$ MeV/$c^2$, a $0^{++}$-state at $1466\pm 6\pm
16$ MeV/$c^2$, and a $0^{++}$-state at $1.7$ GeV/$c^2$ to fit the data
properly.  Due to the large $\rho\pi$-background in $J/\psi$ decays, the
focus was on structures below 2 GeV/$c^2$, but it is worth mentioning
that a peak around 2.1 GeV/$c^2$ was observed; this was the case also
in $\gamma K^+K^-$.  In $\gamma K^+K^-$, the $f_2'(1525)$ and
$f_0(1710)$ were prominent \cite{jpsikk}.

An earlier analysis, using a smaller sample of $(3.90\pm 0.21)\times 10^6$
good quality $\psi(2S)$-events from BESI and using a $\tau^+\tau^-$
background sample at $\sqrt{s}=3.55-3.6$ GeV, showed a clear
$f_2(1270)$ [$BR(\psi(2S)\rightarrow\gamma
f_2(1270))=(2.12\pm0.19\pm0.32)\times10^{-4}$], an $f_0(1710)$ signal
in $\gamma\pi\bar{\pi}$ [$BR(\psi(2S)\rightarrow\gamma
f_0(1710))\times BR(f_0(1710)\rightarrow \pi\bar{\pi})=
(3.01\pm0.41\pm1.24)\times 10^{-5}$], and a clear $f_0(1710)$
[$BR(\psi(2S)\rightarrow\gamma f_0(1710))\times
BR(f_0(1710)\rightarrow K^+K^-)= (3.02\pm0.45\pm 0.66)\times
10^{-5}$], and a less clear $f_2'(1525)$ in \mbox{$\gamma K^+K^-$
  \cite{bes1pseudo}.}


\subsection{Fit shape}
The $0^{++}$ resonances are modeled as constant width Breit Wigner
functions, with $1+\cos^2\theta_\gamma$ angular distributions. The
$2^{++}$ resonances are modeled as Blatt Weisskopf dampened d-wave
shapes (with parameters as in Ref. \cite{bszou} and angular
distributions according to Eq. (\ref{eq:wdist})).  Masses and
widths are taken from the Particle Data Group Compilation \cite{PDG}.
For simplicity any possible nonresonant production and background
component is assumed to follow three-body phase space.  All
resonances are treated as incoherent.  The unbinned meson-meson
invariant mass spectrum is fitted with a log likelihood method, and the
scaled background component from the continuum data
sample and the simulated $\psi(2S)\rightarrow\pi^0\pi^+\pi^-$
background are included in the fit but with the opposite sign.

\subsection{Fit quality}
\label{sec:hypothesistest}
The quality of different log likelihood
fits to the same data set can be compared since the ratio
\begin{equation}
R=-2\ln \left|\frac{{\cal L}(fit_A)}{{\cal L}(fit_B)}\right|
\label{eq:lncomp}
\end{equation}
follows a $\chi^2$-distribution \cite{mark3}.  The number of degrees
of freedom for this $\chi^2$-distribution is given by the number of
free parameters in $fit_B$ minus the number of free parameters in
$fit_A$.  In this way, the difference in log likelihood is used for an
hypothesis test, for instance to determine the necessity of weakly
needed resonances such as the $f_0(1500)$ in $\gamma \pi^+\pi^-$ and
the $f_2'(1525)$ in $\gamma K^+K^-$.  Another measure of the quality
of the overall fit is obtained using a Pearson's $\chi^2$-test.

\section{\boldmath Resonances in $\gamma \pi^+\pi^-$} 
\begin{figure}
\begin{center}
\includegraphics[width=\linewidth]{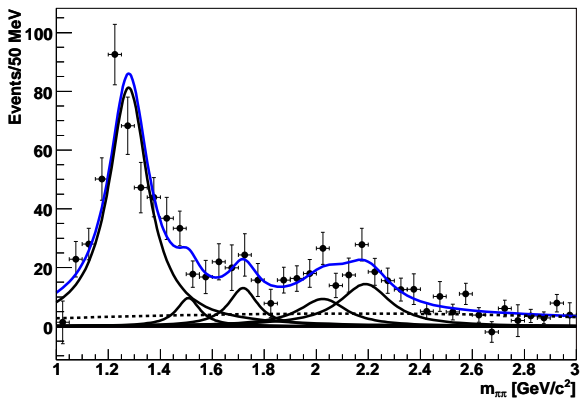}
\caption{Background reduced $m_{\pi^+\pi^-}$ distribution for 
$\gamma\pi^+\pi^-$ events (not efficiency corrected).}
\label{fig:pipifit}
\end{center}
\end{figure}

The fit to the two-pion invariant mass spectrum between 1 and 3
GeV/$c^2$ allows the determination of the number of events and the
product decay branching fractions of $\psi(2S)\rightarrow\gamma
X\rightarrow\gamma \pi^+\pi^-$ where $X=f_2(1270)$, $f_0(1500)$,
$f_0(1710)$, or a high mass component modeled here with $f_4(2050)$
and $f_0(2200)$ (Table \ref{tab:pinumevents}).  The full fit, binned
as in Fig. \ref{fig:pipifit}, has a Pearson's $\chi^2/DOF$ of
$45.7/39$.
\begin{table}
\caption{Number of selected events and product branching fractions in the
$\psi(2S)\rightarrow\gamma X\rightarrow\gamma \pi^+\pi^-$ channel. 
The efficiency at the central resonance mass is determined by Monte Carlo and 
losses due to the limited mass range are taken into account. 
The first uncertainties in the 
branching fractions are statistical and the second systematic.}
\label{tab:pinumevents}
\begin{tabular}{llll}\hline\hline
State ($X$)       & Efficiency  & Events & Branching fraction\\ \hline
$f_2(1270)$  & 13.0\%      & 406$\pm$17 & $(2.2\pm0.1\pm0.2)\times 10^{-4}$ \\
$f_0(1500)$  & 15.0\%      & 31$\pm$14   & $(1.5\pm0.7^{+0.9}_{-0.4})\times 10^{-5}$\\
$f_0(1710)$  & 15.2\%      & 51$\pm$13   & $(2.4\pm0.6^{+0.8}_{-1.1})\times 10^{-5}$\\
$f_4(2050)$  & 15.4\%\footnote{Linear interpolation between $f_0(1710)$ and $f_0(2200)$.} & 61$\pm$20 & $(2.8\pm0.9^{+0.8}_{-0.6})\times 10^{-5}$\\
$f_0(2200)$  & 15.6\%      & 99$\pm$22  & $(4.6\pm1.0^{+4.5}_{-0.9})\times 10^{-5}$  \\ \hline\hline
\end{tabular}
\end{table}
Note that the invariant mass peak near the $f_2(1270)$ is lower in
this data set than the central value given by the Particle Data Group
\cite{PDG} (Fig. \ref{fig:pipifit}), and the branching fraction into
$f_2(1270)$ is almost twice as large as in the corresponding BESI
measurement \cite{bes1pseudo}. The branching fraction into $f_0(1710)$
agrees with the BESI measurement, and for $f_0(1500)$ a branching
fraction was not given by BESI.  We use the $\chi^2$-hypothesis test
(Section \ref{sec:hypothesistest}) to check whether the inclusion of
the $f_0(1500)$ improves the fit enough to motivate the one extra fit
parameter.  Removing the $f_0(1500)$ here changed the log likelihood
such that $R=6.2$ (Eq. \ref{eq:lncomp}). In a $\chi^2$-distribution
with one degree of freedom, the probability to be above 6.2 is 1.28\%,
and, since 1.28\% $ < $ 5\%, the $f_0(1500)$ is accepted at the 5\% level.
The high mass region, which is fitted with $f_4(2050)$ and
$f_0(2200)$, can also be well fitted by alternative combinations of
resonances; combinations with $f_2(1810)$, $f_0(2020)$, $f_2(2150)$,
$f_4(2050)$, $f_0(2200)$, and a free mass and free width state have
been tested.  The selected fit is the best, but since other
alternatives are not ruled out they are used to estimate the
systematic uncertainty to the other branching fractions.  In one version
of the fit, the event excess in the region between two-pion threshold
and 800 MeV/$c^2$ (Fig. \ref{fig:pi}) is fitted with an additional
$\sigma$-shape. This is used to give a measure of the systematic
uncertainty introduced by neglecting the $\sigma$.

\section{\boldmath Resonances in $\gamma K^+K^-$}
\label{sec:kk}
\begin{table}
\caption{Number of selected events and product branching fractions in the
$\psi(2S)\rightarrow\gamma X\rightarrow\gamma K^+K^-$ channel. 
The efficiency at the central resonance mass is determined by Monte Carlo 
and losses due to the limited mass range 
are taken into account. 
The first uncertainties in the 
branching fractions are statistical and the second systematic.}
\label{tab:knumevents}
\begin{tabular}{llll}\hline\hline
State ($X$)        & Efficiency  & Events & Branching fraction\\ \hline
$f_2(1270)$  & 14.4\%      & 39$\pm$12  & $(1.9\pm0.6^{+1.0}_{-0.6})\times 10^{-5}$ \\
$f'_2(1525)$  & 15.6\%     & 15$\pm$10   & $(6.9\pm4.4^{+4.1}_{-2.1})\times 10^{-6}$\\
$f_0(1710)$  & 16.3\%      & 70$\pm$14  & $(3.1\pm0.6^{+1.1}_{-0.7})\times 10^{-5}$\\ 
\hline\hline
\end{tabular}
\end{table}

\begin{figure}
\begin{center}
\includegraphics[width=\linewidth]{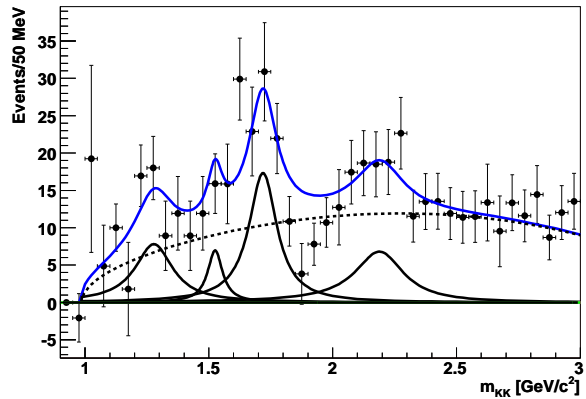}
\caption{Background reduced $m_{K^+K^-}$ distribution for 
$\gamma K^+K^-$ events (not efficiency corrected).}
\label{fig:kkfit}
\end{center}
\end{figure}

Fitting the $m_{K^+K^-}$ distribution between threshold and 3
GeV/$c^2$, the product branching fractions for $\psi(2S)\rightarrow\gamma
X\rightarrow\gamma K^+K^-$, where $X=f_2(1270)$, $f_2'(1525)$, and
$f_0(1710)$, are obtained with a Pearson's $\chi^2/DOF$ of $37.4/39$
(Table \ref{tab:knumevents} and Fig. \ref{fig:kkfit}).  The high mass
region is fitted with a three-body phase space background component
and a combination of resonances: $f_4(2050)$, $f_0(2100)$,
$f_2(2150)$, $f_0(2200)$, $f_2(2300)$, $f_2(2350)$, and $f_6(2510)$.
The fit with $f_0(2200)$ and a phase space background is used to
calculate the branching fractions in Table \ref{tab:knumevents} (Fig.
\ref{fig:kkfit}), and the other fits are used to estimate the
systematic uncertainty.  The $f_2(1270)$ branching fraction is a
factor four larger than the BESI measurement, but consistent with the
measurement in $\gamma\pi^+\pi^-$ within the (large) uncertainties.
The branching fraction into $f_0(1710)$ agrees with the BESI
measurement \cite{bes1pseudo}, and the 
$f_2'(1525)$ was part of the BESI fit \cite{bes1pseudo}, but a
branching fraction was never given.  Removing the $f_2'(1525)$ here
changed the log likelihood such that $R=3.8$ (Eq. \ref{eq:lncomp}),
which corresponds to a statistical significance of less than 2$\sigma$. 

\section{Uncertainties}
The statistical uncertainty includes the uncertainty in the number of
events from data, continuum processes, and the small statistical
uncertainty from the Monte Carlo simulation of
$\psi(2S)\rightarrow\pi^+\pi^-\pi^0$, which is propagated into the
uncertainty of the fitted branching fraction.  Systematic uncertainties
are inevitably introduced due to differences between simulation and
reality; in this case these are associated with MDC tracking, the
kinematic fit, and particle identification (Table \ref{tab:syserr}).
The detection efficiency for any state also depends on the angular
distribution of the decay. For $J^{PC}=2^{++}$ states, the angular
distribution is determined up to two real parameters, $x$ and $y$.  In
section \ref{sec:xyfit} these are fitted to $\psi(2S)\rightarrow\gamma
f_2(1270)\rightarrow\gamma\pi^+\pi^-$ data, and two different sign
solutions are found.
The overall detection efficiency of the two cases have a relative
difference of 2.3\% in $\gamma\pi^+\pi^-$ and 2.8\% in $\gamma
K^+K^-$; this uncertainty is added quadratically into the overall
systematic uncertainty for the $J^{PC}=2^{++}$ and $4^{++}$ states.

\begin{table}
\caption{Sources of systematic uncertainty to the 
branching fractions. Fit uncertainties are not included.}
\label{tab:syserr}
\begin{tabular}{ll}\hline\hline
Source of uncertainty                 &   \\ 
\hline
Number of $\psi(2S)$ \cite{numberofevents} & 4\%      \\
\hline
Charged tracks \cite{psipradiative} & 4\%    \\
Photon ID  \cite{syserror}   & 2\%    \\
Trigger \cite{psipradiative}        & $<$0.5\%  \\
Kinematic fit \cite{kinematicfit} & 6\%\\
\hline
Eff. variation with (x,y) $m_{\pi^+\pi^-}$    & 2.3\%  \\
Eff. variation with (x,y) $m_{K^+K^-}$        & 2.8\%  \\
\hline\hline
\end{tabular}
\end{table}

The main source of systematic uncertainty is the choice of fit
function, where the modeling of the background plays a major part.
The $\psi(2S)\rightarrow\pi^+\pi^-\pi^0$ also contributes to the
systematic uncertainty, since the branching fraction from the BESII measurement
\cite{bes3pi} has a relative
uncertainty of 14.5\%, which for
$\psi(2S)\rightarrow\gamma\pi^+\pi^-$ corresponds to $\sim$15 events
in the region between 1 and 3 GeV/$c^2$.  In the fit to
$m_{\pi^+\pi^-}$, a few weak background sources are ignored; these
add up to $\sim 5$ events in the region between 1
GeV/$c^2<m_{\pi\pi}<2.5$ GeV/$c^2$.  There could also be a
contribution from unknown background and non-resonant production. The
uncertainty introduced from these sources is investigated by adding a
three-body phase space component to the fit, which converges to
$\sim$150 events between $m_{\pi\pi}=$ 1 and 3 GeV/$c^2$. This is a
component which is necessary to fit the data,
$P_{\chi^2}(64,1)<0.0001$.  The corresponding background component in
$m_{K^+K^-}$ is more prominent (Fig. \ref{fig:kkfit}), and the
branching fractions of some possible background sources are not yet
measured.  To investigate how the uncertainty in the background
estimate influences the branching fractions in $\gamma\pi^+\pi^-$ and
$\gamma K^+K^-$, the best fit is compared to a fit with a background
fraction which is 50\% smaller and 50\% larger.  The overall
uncertainty introduced by the fit function is investigated by taking
the minimum, and the maximum, value of the branching fraction for each
specific resonance, with different resonance combinations and
different size background components, and comparing this with the best
fit. After adding the uncertainties in quadrature, we obtain
the systematic errors in Tables \ref{tab:pinumevents} and
\ref{tab:knumevents}.


\section{\boldmath Angular distribution of $\psi(2S)\rightarrow\gamma f_2(1270)\rightarrow\gamma\pi^+\pi^-$}
\label{sec:xyfit}
The radiative decay of a vector state [such as the $J/\psi$
or $\psi(2S)$] into a tensor state [such as the light quark tensor
$f_2(1270)$], which decays further into two pseudoscalars, has an
angular distribution, which is determined by two real parameters:
$x=A_1/A_0$ and $y=A_2/A_0$ \cite{Kabir, Edwards}. Here $A_0$, $A_1$,
and $A_2$ are the helicity amplitudes for projection zero, one and
two, respectively.

\begin{widetext}
\begin{eqnarray}
\label{eq:wdist}
W(\theta_\gamma,\theta_M, \phi_M)&=& 
  3x^2\sin^2\theta_\gamma\sin^22\theta_M
 +(1+\cos^2\theta_\gamma)\left[(3\cos^2\theta_M-1)^2+
    \frac{3}{2}y^2\sin^4\theta_M\right] \\ \nonumber
+\sqrt{3}x\sin 2\theta_\gamma \sin 2\theta_M & &\left[3\cos^2\theta_M-1-
    \frac{\sqrt{6}}{2}y\sin^2\theta_M \right]\cos\phi_M
+ \sqrt{6}y\sin^2\theta_\gamma \sin^2\theta_M (3\cos^2\theta_M-1)\cos 2\phi_M.
\end{eqnarray}
\end{widetext}

In the BESII cylindrical geometry, the $z$-axis is defined by the
beam, $\theta_\gamma$ is the polar angle of the direct photon, and
$\theta_M$ is the polar angle of one of the pions with respect to the
$\gamma$-direction in the $f_2(1270)$ rest frame.  The last parameter,
$\phi_M$, is the azimuthal angle of the same pion in the $f_2(1270)$
rest frame. By convention $\phi=0$ lies in the plane defined by the
electron beam and the photon in the $f_2(1270)$ rest frame.
Predictions for $x$, $y$ have been based on perturbative QCD
\cite{Krammer}, helicity two suppression \cite{Close1983}, or, for
instance, assumptions about the intermediate resonance being a
glueball \cite{Ward} (Table \ref{tab:xytable}).

\begin{table}
\caption{Helicities in different decay models and in measurement of
$J/\psi\rightarrow\gamma f_2(1270)\rightarrow\gamma\pi^+\pi^-$. $x$ and $y$ are defined 
in the text.}
\label{tab:xytable}
\begin{tabular}{lcc}\hline\hline
Model      & x & y \\ \hline
E1, low energy photon, pQCD \cite{Krammer} & $\sqrt{3}$ & $\sqrt{6}$ \\
Krammer, high energy photon \cite{Krammer} & $0$ & $0$ \\
Krammer at $\psi(2S)\rightarrow\gamma f_2(1270)$ \cite{Krammer} & $0.68$ & $0.45$ \\ 
Close \cite{Close1983} & $\sqrt{3}/2$ & $0$ \\
Ward glueball \cite{Ward} & $-0.85$ & $-1.0$ \\
$J/\psi$ BESII \cite{jpsipwa}  & $0.89\pm0.02$     & $0.46\pm0.02$ \\
$J/\psi$ Crystal Ball \cite{Edwards} & $0.88\pm0.13$     & $0.04\pm0.19$ \\
$J/\psi$ MARK III \cite{mark3} & $0.96\pm0.07$     & $0.06\pm0.08$ \\ 
\hline\hline
\end{tabular}
\end{table}

The $W$-distribution (Eq. \ref{eq:wdist}) depends on the parameter
$x$, but projections of the $W$-distribution onto the
$\cos\theta_\gamma$ and the $\phi_M$ axes become independent of $x$,
and the projection onto $\cos\theta_M$ is symmetric in $x$
\cite{Lundborg}. Therefore a three-dimensional analysis of many events
is needed in order to determine the size and the sign of $x$
\cite{mark3, Edwards, Pluto}.  The 172 events with
$|m_{\pi^+\pi^-}-m_{f_2(1270)}|<0.5 \Gamma_{f_2(1270)}$ (mass and width
from Ref. \cite{PDG}) are fitted with a log likelihood fit. Continuum
(15.4 events after scaling) and $\psi(2S)\rightarrow\pi^+\pi^-\pi^0$
(5.6 events after scaling) backgrounds are treated as scaled opposite
sign contributions. The detection efficiency is determined using Monte
Carlo modeling. Here, the cylindrical structure and the event
selection requirements give a detection efficiency with dips at the
end-caps in $\cos\theta_\gamma$ (as in MARKIII \cite{mark3}).  The
detection efficiency is relatively isotropic in $\cos\theta_M$ but
varies strongly with $\phi_M$, where minima are obtained at
$\phi_M=0$, $\pi$, and $2\pi$; in these angular regions mesons can be
close to the beam directions. Since both $\theta_\gamma$ and $\phi_M$
suffer from large acceptance correction effects, the $\theta_M$
distribution is the most powerful analyzer.

An angular fit to the three-dimensional $\psi(2S)\rightarrow\gamma
f_2(1270)\rightarrow\gamma \pi^+\pi^-$ distribution gives two
solutions for $x$ and $y$ (Table \ref{tab:xyfit}).  The component with
helicity two, $y$, is well determined whereas the component with
helicity one, $x$, is essentially undetermined, mainly due to
acceptance losses.  The 1$\sigma$ uncertainty in the fits is stated
next to the fit result in Table \ref{tab:xyfit}; these are equivalent
to the statistical uncertainties from data and background. 

Systematic uncertainties are listed in Table \ref{tab:wsyserr}.  The
effects of insufficiently understood acceptance was investigated in
Ref.  \cite{shixinpaper} by comparing the fit to
$\psi(2S)\rightarrow\gamma\chi_{c0}$ data with the known angular
distribution, revealing a large systematic uncertainty in $x$.  The
effect of $\psi(2S)\rightarrow\gamma\pi^+\pi^-$ background not
resonating in $f_2(1270)$ is investigated by varying the invariant
mass region, and effects of misidentified contributions within the data
set are investigated by comparing the fit with and without continuum
and $\psi(2S)\rightarrow\pi^+\pi^-\pi^0$ contributions.  Muon
contamination is automatically part of the opposite sign continuum
sample, and, since simulation showed no
$\psi(2S)\rightarrow\mu^+\mu^-$-contribution and less than a 1\%
$e^+e^-\rightarrow\mu^+\mu^-$ contribution within the region, these are
assumed to be negligible.  Misidentification of $\gamma K^+K^-$
events is estimated to give a contribution of 0.6 events within the
region and systematic uncertainties from the generator input-output
are negligible.  When the correlation between the uncertainties in $x$
and $y$ is unknown the maximum is assumed, $i.e.$ 1 and -1 for the
positive and the negative solution respectively, for nonresonant events
and misidentified background.  The overall correlation $\rho$ between
systematic uncertainties $\sigma$ is calculated as
\begin{equation}
\rho=\sum_i \frac{\rho_i\sigma_{xi}\sigma_{yi}}{\sigma_x\sigma_y}
\end{equation} 
where $i$ runs over all the systematic uncertainties \cite{shixinpaper}.

\begin{table}
\caption{The relative fraction of helicity one $x$ and two $y$, to helicity zero in
$\psi(2S)\rightarrow\gamma f_2(1270)\rightarrow\gamma\pi^+\pi^-$. With the given number of events
a preferred solution cannot be chosen. The first uncertainty is the 1$\sigma$-deviation in
the fit, $i.e.$ the statistical uncertainty, the second is systematic.}
\label{tab:xyfit}
\begin{tabular}{ll}\hline\hline
Positive solution         & Negative solution\\ \hline
$x=0.20\pm0.09\pm0.25$    & $x=-0.26\pm0.09\pm0.24$  \\
$y=-0.26\pm0.08\pm0.05$   & $y=-0.25\pm0.09\pm0.06$ \\
$\rho_{stat}=0.53$        & $\rho_{stat}=-0.43$ \\
$\rho_{sys}=0.44$         & $\rho_{sys}=-0.41$ \\
\hline \hline
\end{tabular}
\end{table}

\begin{table}
\begin{center}
\caption{Systematic uncertainties in the angular distribution fit.}
\label{tab:wsyserr}
\begin{tabular}{lll}\hline\hline
Source of uncertainty & Positive  & Negative  \\ \hline
\begin{tabular}{l}
\\
Angular acceptance \\ 
Non-resonant events \\ 
(Fit region variation)\\
Misidentified background\\
(With/without bg red.) \\  
Overall     \\      
\end{tabular} 
&
\begin{tabular}{lll}
$\sigma_x$  & $\sigma_y$  & $\rho$ \\ 
0.18        & 0.05        & 0.24   \\
0.14     & 0.02        & 1      \\ 
     &             &        \\ 
0.11     & 0.01        & 1      \\
   &             &        \\

 0.25     & 0.05        & 0.44   \\

\end{tabular} 
&
\begin{tabular}{lll}
$\sigma_x$  & $\sigma_y$  & $\rho$ \\ 
 0.18        & 0.05        & -0.24   \\
0.12     & 0.03        & -1      \\ 
     &             &        \\ 
 0.10     & 0.00           & -1      \\
     &             &        \\
 0.24     & 0.06        & -0.41   \\

\end{tabular}\\
\hline\hline
\end{tabular}
\end{center}
\end{table}

\section{Conclusion and discussion}

\begin{table}
\begin{center}
\caption{The ratio $BR(\psi(2S)\rightarrow\gamma
  X)/BR(J/\psi\rightarrow\gamma X)$ for comparison with the 12 \% Rule.
  The $J/\psi$ branching fractions in the middle column are taken
  from \cite{jpsipwa, jpsikk}, and those in the right most column are
  taken from the Particle Data Group \cite{PDG}.}
\label{tab:12procent}
\begin{tabular}{lll}\hline \hline
$X$                              & BES \cite{jpsipwa, jpsikk}  & PDG \cite{PDG} \\ \hline
$f_2(1270)\rightarrow\pi^+\pi^-$ & 24$\pm$6\%       &  28$\pm$6 \% \\
$f_0(1500)\rightarrow\pi^+\pi^-$ & 22$\pm$27\%      &  $>5\pm$4\% \\
$f_0(1710)\rightarrow\pi^+\pi^-$ & 9$\pm$5\%        &  14$\pm$18\% \\
\hline
$f_2(1270)\rightarrow K^+K^-$    & -                &  60$\pm$45 \% \\
$f_2'(1525)\rightarrow K^+K^-$   & 4$\pm$6\%        &  3$\pm$4 \% \\
$f_0(1710)\rightarrow K^+K^-$    & 6$\pm$5\%        &  7$\pm$4 \% \\
\hline \hline
\end{tabular}
\end{center}
\end{table}

The branching fraction of $\psi(2S)\rightarrow \gamma f_0(1710)$ is
measured in both the $\pi^+\pi^-$ and the $K^+K^-$ decays. Both
results are close to the one previous measurement from BES
\cite{bes1pseudo}.  The ratio between $f_0(1710)\rightarrow
\pi^+\pi^-$ and $f_0(1710)\rightarrow K^+K^-$ is found to be
(77$\pm$72)\%, with a large uncertainty mainly from the $K^+K^-$
measurement.  The potential glueball, the $f_0(1500)$, is accepted at
the 5\%-level in $\gamma\pi^+\pi^-$.  It has been seen in
$J/\psi\rightarrow\gamma\pi^+\pi^-$ \cite{jpsipwa}, but this is a
first measurement in $\psi(2S)$ radiative decays. 
The $f_2'(1525)$ is at the verge of being accepted at
the 5\%-level in $\gamma K^+K^-$. The $f_2'(1525)$ has a strong signal
in $J/\psi\rightarrow\gamma K^+K^-$ \cite{jpsikk} and was part of the
fit in the BESI $\psi(2S)\rightarrow\gamma K^+K^-$ analysis
\cite{bes1pseudo}, but a branching fraction was never given.  The region
above 2 GeV/$c^2$ has an enhancement both in $\pi^+\pi^-$ and
$K^+K^-$.  In $\pi^+\pi^-$, the region is fitted with two resonances:
the $f_4(2050)$ and the $f_0(2200)$.  Both of these have been detected
before in pseudoscalar final states in radiative $J/\psi$-decays
\cite{PDG}. In $K^+K^-$, the same region has a rather uniform
distribution with an enhancement around the $f_0(2200)$. The invariant
mass distribution from $\psi(2S)$ decays is similar to the previously
measured $J/\psi$ case but due to the limited number of events here, a
partial wave analysis is not feasible.  One problem when comparing the
$J/\psi$ and the $\psi(2S)$ fit in $m_{\pi^+\pi^-}$ is that in
$J/\psi$ there is destructive interference between the $f_0(2020)$ and
the $f_0(1710)$ causing a dip at $\sim$1.8 GeV/$c^2$.  There is also
destructive interference around 1.5 GeV/$c^2$, which shifts the peak
of the $f_0(1500)$ to lower values than the pole position
\cite{jpsipwa}. This pattern is not reproduced if we, as in this
analysis, assume incoherence.  The ratio $BR(\psi(2S)\rightarrow\gamma
X)/BR(J/\psi\rightarrow\gamma X)$ is presented in Table
\ref{tab:12procent} for a few final states $X$ for comparison with the
12\%-rule.

Can we use the branching fractions to understand what goes on during the
decay process?  In an old prediction by Lipkin \cite{Lipkin},
different pQCD and pQED $\psi\rightarrow \gamma 2^{++}$ decay diagrams
are compared under the assumption of isospin symmetry.
\begin{figure}
\includegraphics[width=0.8\linewidth]{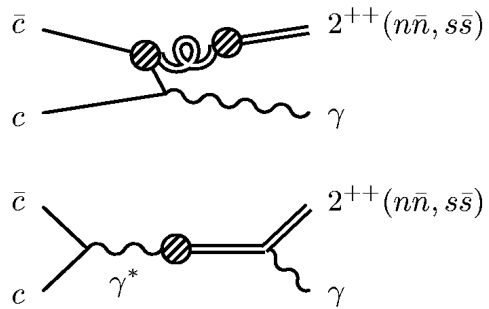}
\caption{Using perturbative QCD and QED, the process $\psi\rightarrow\gamma 2^{++}(q\bar{q})$ 
has an affinity for $s\bar{s}$ and $n\bar{n}$, which depends on the preferred 
diagram \cite{Lipkin}. Lipkin predicted a ratio between 
$s\bar{s}$ and $n\bar{n}$ of 50\% for the upper diagram and 8\% for the lower, 
without taking phase space into account.}
\label{fig:TensorProcess}
\end{figure}
The upper diagram in Fig. \ref{fig:TensorProcess} gives the ratio
\begin{equation}
R=\frac{\Gamma(\psi(2S)\rightarrow \gamma f_2'(1525))}{\Gamma (\psi(2S)\rightarrow\gamma f_2(1270))}=0.5
\end{equation}
and the lower gives $R\approx 8\%$. 
The analysis presented here gives a ratio of $\approx4\%$ whereas 
corresponding measurement in $J/\psi$ gave $R=26$\%.
From another pQCD diagram with photon radiation and subsequent hadron formation, Close
predicted \cite{Close1983}
\begin{equation}
\frac{BR(\psi\rightarrow\gamma 0^+)}{BR(\psi\rightarrow\gamma 2^+)}=\frac{2}{7}
\end{equation}
under the assumption of quark-antiquark pair formation with $L_z=1$ and identical coupling
between glue and scalar and glue and tensor. 
This can unfortunately not be compared to the
present data since $BR(f_0(1710)\rightarrow K^+K^-)$ is unknown.

\begin{acknowledgments}
The BES collaboration thanks the staff of BEPC and computing center for 
their hard efforts. 
This work is supported in part by the National Natural Science Foundation of 
China under contracts Nos. 10491300, 10225524, 10225525, 10425523, the Chinese Academy of Sciences under contract No. KJ 95T-03, the 100 Talents Program of CAS under Contract Nos. U-11, U-24, U-25, and the Knowledge Innovation Project of CAS under Contract Nos. U-602, U-34 (IHEP), the National Science Foundation of China under Contract No. 10225522 (Tsinghua University), the Department of Energy under Contract No. DE-FG02-04ER41291 (U. Hawaii), 
the Royal Physiographic Society in Lund Sweden, 
the Uppsala University Graduate School $\Delta U$, the Sederholm
Foundation, and the Gertrud Thelin foundation at Uppsala University.
\end{acknowledgments}

 
\end{document}